\documentclass[a4paper]{PoS}

\usepackage{amssymb,amsmath,amsfonts}

\newcommand{\bea}{\begin{eqnarray}}
\newcommand{\ena}{\end{eqnarray}}
\newcommand{\nn}{\nonumber\\}
\newcommand{\be}{\begin{equation}}
\newcommand{\en}{\end{equation}}

\newcommand{\ed}{\end{document}}

\newcommand{\la}{\langle}
\newcommand{\ra}{\rangle}

\newcommand{\slp}{p\kern-5pt/}

\title{The invariant and helicity amplitudes in the transitions 
$\Lambda_b\to\Lambda^\ast(\frac12^\pm,\frac32^\pm)+J/\psi$}

\ShortTitle{The invariant and helicity amplitudes in the transitions
$\Lambda_b\to\Lambda^\ast(\frac12^\pm,\frac32^\pm)+J/\psi$.}

\author{\speaker{Mikhail Ivanov}\thanks{Talk based on the work
in collaboration with
T.~Gutsche, J.~G.~K\"orner, V.~E.~Lyubovitskij, 
V.~V.~Lyubushkin, P.~Santorelli and C.~T.~Tran.
}\\
        JINR, Dubna\\
        E-mail: \email{ivanovm@theor.jinr.ru}}

\abstract{We present results for the invariant and helicity amplitudes 
in the transitions  $\Lambda_b\to\Lambda^{(\ast)}(J^P)+J/\psi$ where
the $\Lambda^{(\ast)}(J^P)$ are  $\Lambda(sud)$-type ground and excited states
with $J^P$ quantum numbers $J^P=\frac12^{\pm},\frac32^{\pm}$. 
The calculations are performed in the framework of a covariant confined quark 
model. This analysis is important for the identification of
the hidden charm pentaquark states $P_c^+(4380)$ and $P_c^+(4450)$ which
were discovered by the LHCb Coll. 
We also  discuss the possible New Physics effects in the exclusive 
decays $\bar{B}^0 \to D^{(\ast)} \tau^- \bar{\nu}_{\tau}$. 
We extend the Standard Model by taking into account right-handed vector 
(axial), left- and right-handed (pseudo)scalar, and tensor current 
contributions. The $\bar{B}^0 \to D^{(\ast)}$ transition form factors are 
calculated in the full kinematic $q^2$ range by employing a covariant quark 
model.
  }

\FullConference{The European Physical Society Conference on High Energy 
Physics\\
                 5-12 July 2017\\
                 Venice, Italy}

\begin{document}

\section{Introduction}

\label{sec:intro} 
In the last few years, the semileptonic decays
$\bar{B}^0 \to D^{(\ast)} \tau^- \bar{\nu}_{\tau}$ have been widely discussed in
the literature as candidates for testing the Standard Model (SM) and 
searching for possible new physics (NP) in charged-current interactions. 
At $B$ factories, the Belle and \textit{BABAR} Collaborations  have been 
continuously updating their measurements with better precision based on 
electron-positron colliders. Recently, the LHCb Collaboration has also entered 
the game with data taken at the LHC hadron collider. The three groups have 
reported measurements of the ratios in 
Refs.~\cite{Lees:2012xj,Huschle:2015rga,Aaij:2015yra,Sato:2016svk,
Abdesselam:2016xqt}. These measurements provide the average ratios
\be
R(D)|_{\rm expt} = 0.397 \pm 0.049 ,
\qquad
R(D^\ast)|_{\rm expt} = 0.308 \pm 0.017 ,
\label{eq:RD-expt}
\en
which exceed the SM  expectations given in Refs.~\cite{Na:2015kha,Fajfer:2012vx}
\be
R(D)|_{\rm SM} = 0.300 \pm 0.008 ,
\qquad
R(D^\ast)|_{\rm SM} = 0.252 \pm 0.003 ,
\label{eq:RD-SM}
\en
by 1.9~$\sigma$ and 3.3~$\sigma$, respectively.
The excess of $R(D^{(\ast)})$ over SM predictions has attracted a great deal
of attention in the particle physics community and has led to many theoretical
studies looking for NP explanations.

In the paper \cite{Ivanov:2016qtw} we included NP operators in the effective 
Hamiltonian and investigated their effects on physical observables of 
the decays 
$\bar{B}^0 \to D^{(\ast)} \ell^-\bar\nu_{\ell}$. We defined a full set of form 
factors corresponding to SM+NP operators and calculated them by employing 
the covariant confined quark model (CCQM). 
In the CCQM the transition form factors can be determined in the full
range of  momentum transfer, making the calculations straightforward without 
any extrapolation. This provides an opportunity to investigate NP operators in 
a self-consistent manner, and independently from the HQET. We first constrain 
the NP operators using experimental data, then analyze their effects on various
observables including the ratios of branching fractions, 
the forward-backward asymmetries, and a set of polarization observables. 
We also derive the fourfold angular distribution for the cascade decay 
$\bar {B}^0\to D^{\ast\,+}(\to D^0\pi^+)\tau^-\bar\nu_\tau$ to analyze 
the polarization of the $D^{\ast}$ meson in the presence of NP by using the 
traditional helicity amplitudes.

Recently the LHCb Collaboration has performed an angular
analysis of the decay $\Lambda_b \to \Lambda^{(*)} + J/\psi$,
where the $\Lambda_{b}$'s are produced in $pp$ collisions
at $\sqrt{s} = 7$ TeV at the LHC (CERN)~\cite{Aaij:2013oxa}.
They reported on the measurement of the relative magnitude of the helicity
amplitudes in the decay $\Lambda_b \to \Lambda^{(*)} + J/\psi$ by a fit to
several asymmetry parameters in the cascade decay distribution
$\Lambda_b \to \Lambda(\to p\pi^-) + J/\psi (\to \ell^{+}\ell^{-})$ and
$\Lambda_b \to \Lambda^\ast(\to pK^-) + J/\psi (\to \ell^{+}\ell^{-})$. 
In the paper~\cite{Gutsche:2013oea}  we have performed a detailed
analysis of the decay process $\Lambda_b \to \Lambda + J/\psi$ within the CCQM.
We have worked out two variants of the threefold joint angular decay 
distributions in the cascade decay
$\Lambda_b\to \Lambda(\to p\pi^-)\,+\,J/\psi(\to\ell^+\ell^-)$ for polarized
and unpolarized $\Lambda_{b}$ decays. We have further listed results on
helicity amplitudes which determine the rate and the asymmetry
parameters in the decay processes
$\Lambda_b \to \Lambda(\to p\pi^-)\,+\,J/\psi$ and
$\Lambda_b \to \Lambda(\to p\pi^-)\,+\,\psi(2S)$.

In the paper~\cite{Gutsche:2017wag} we have calculated the corresponding 
invariant and helicity amplitudes in
the transitions $\Lambda_b~\to~\Lambda^{(\ast)}(J^P)~+~J/\psi$ where the
$\Lambda^{(\ast)}(J^P)$ are $\Lambda$-type $(sud)$ ground and excited states with
$J^P$ quantum numbers $J^P=\frac12^{\pm},\frac32^{\pm}$. 
We found that the values of the helicity amplitudes for the
$\Lambda_b \to \Lambda^\ast(1520,\,\frac32^-),\Lambda^\ast(1890,\,\frac32^+)$
transitions are suppressed compared with those for the
transitions to the ground state
$\Lambda(1116,\,\frac12^+)$ also calculated in~\cite{Gutsche:2013oea}
and the excited state $\Lambda^\ast(1405,\,\frac12^-)$.
This analysis is important for the identification of
the hidden charm pentaquark states $P_c^+(4380)$ and $P_c^+(4450)$ since the 
cascade decay 
$\Lambda_b~\to~\Lambda^\ast(\frac12^-,\frac32^\pm)(~\to~pK^-)~+~J/\psi$
involves the same final  states as the decay
$\Lambda_b^0~\to~P_c^+(~\to~p~J/\psi)~+~K^- $. 

\section{Extended Effective Hamiltonian for 
$B\to D^{(\ast)}\tau\bar\nu_\tau$ decay}

We extend the SM effective Hamiltonian for the quark-level transition 
$b \to c \tau^- \bar{\nu}_{\tau}$ by including new operators: 
\bea
{\mathcal H}_{eff} &=&
2\sqrt {2}G_F V_{cb}[(1+V_L)\mathcal{O}_{V_L}+V_R\mathcal{O}_{V_R}
+S_L\mathcal{O}_{S_L}+S_R\mathcal{O}_{S_R} +T_L\mathcal{O}_{T_L}],
\label{eq:Heff}\\[2ex]
\mathcal{O}_{V_L} &=& 
\left(\bar{c}\gamma^{\mu}P_Lb\right)
\left(\bar{\tau}\gamma_{\mu}P_L\nu_{\tau}\right), \qquad
\mathcal{O}_{V_R} =
\left(\bar{c}\gamma^{\mu}P_Rb\right)
\left(\bar{\tau}\gamma_{\mu}P_L\nu_{\tau}\right),
\nn
\mathcal{O}_{S_L} &=&
\left(\bar{c}P_Lb\right)\left(\bar{\tau}P_L\nu_{\tau}\right),\qquad
\mathcal{O}_{S_R} =
\left(\bar{c}P_Rb\right)\left(\bar{\tau}P_L\nu_{\tau}\right),
\qquad
\mathcal{O}_{T_L} =
\left(\bar{c}\sigma^{\mu\nu}P_Lb\right)
\left(\bar{\tau}\sigma_{\mu\nu}P_L\nu_{\tau}\right).
\nonumber
\ena
Here, $\sigma_{\mu\nu}=i\left[\gamma_{\mu},\gamma_{\nu}\right]/2$, 
$P_{L,R}=(1\mp\gamma_5)/2$ are the left and right projection operators, and 
$V_{L,R}$, $S_{L,R}$, and $T_L$ are the complex Wilson coefficients governing 
the NP contributions. In the SM one has $V_{L,R}=S_{L,R}=T_L=0$. We assume that 
NP only affects leptons of the third generation.

The form factors which appear in the matrix elements including the NP
operators are calculated in the framework of the CCQM. The model parameters, 
namely, the hadron size parameter $\Lambda$, the constituent quark masses 
$m_{q_i}$, and the universal infrared cutoff parameter $\lambda$, are 
determined by fitting calculated quantities 
of a multitude of basic processes to available experimental data or 
lattice simulations.
It is important to note that within the SM (without any NP operators) our 
model calculation yields 
$R(D)=0.267$ and $R(D^\ast)=0.238$~\cite{Ivanov:2015tru},
which are consistent with other SM predictions given in 
Refs.~\cite{Na:2015kha, Fajfer:2012vx} within $10\%$.

\begin{figure}[htbp]
\begin{tabular}{lr}
\includegraphics[scale=0.45]{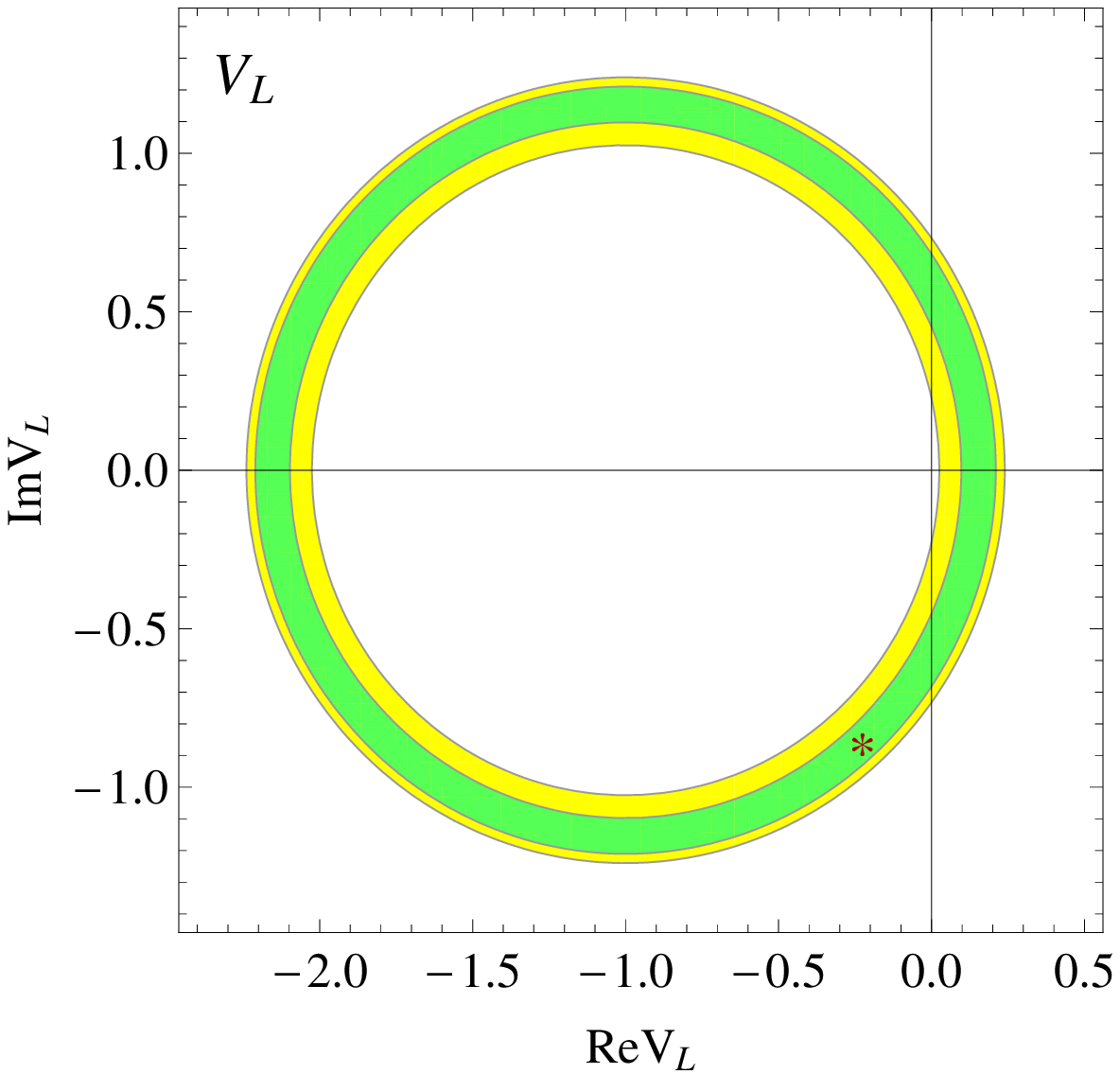}
& 
\includegraphics[scale=0.45]{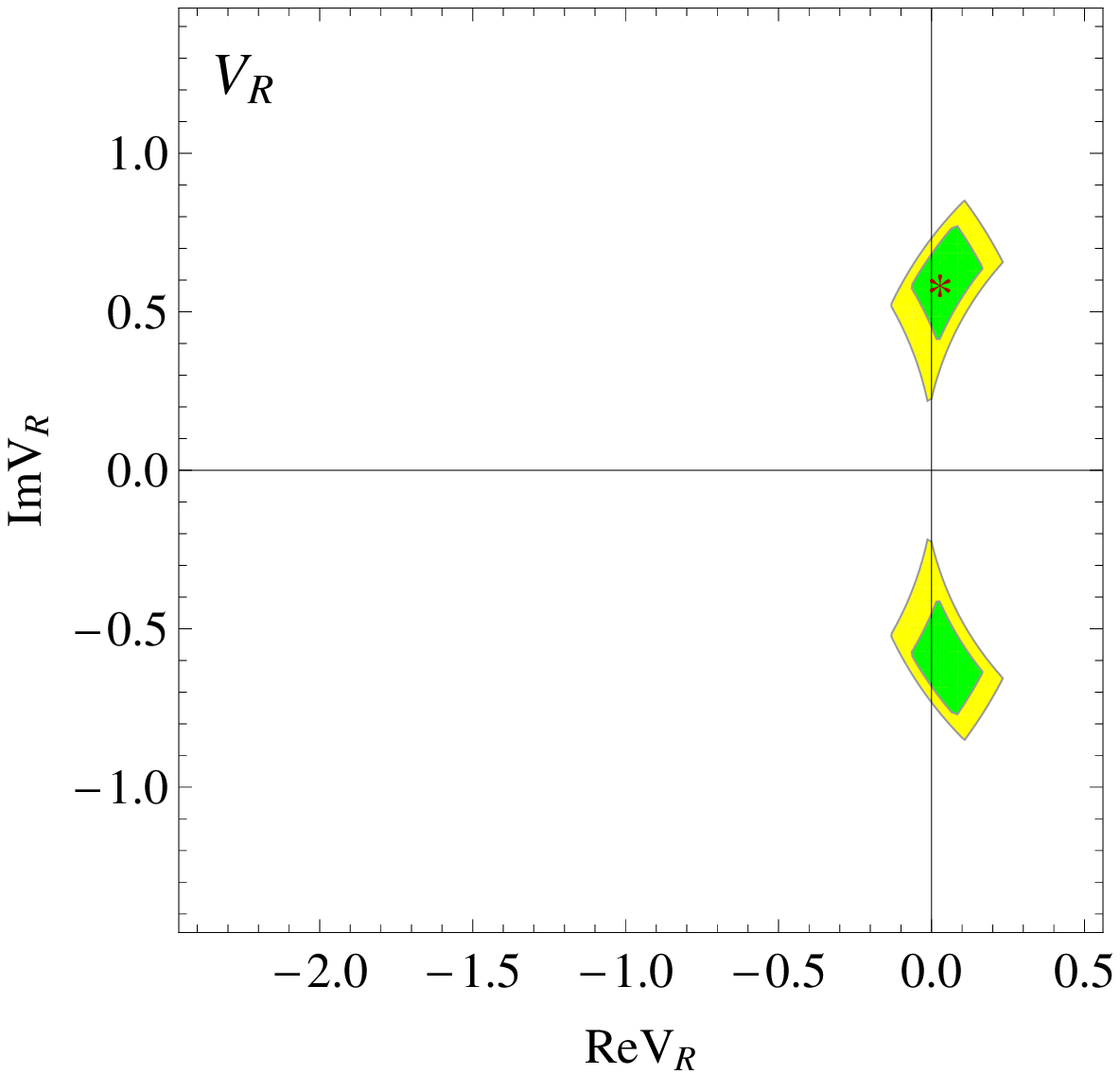}\\
\includegraphics[scale=0.45]{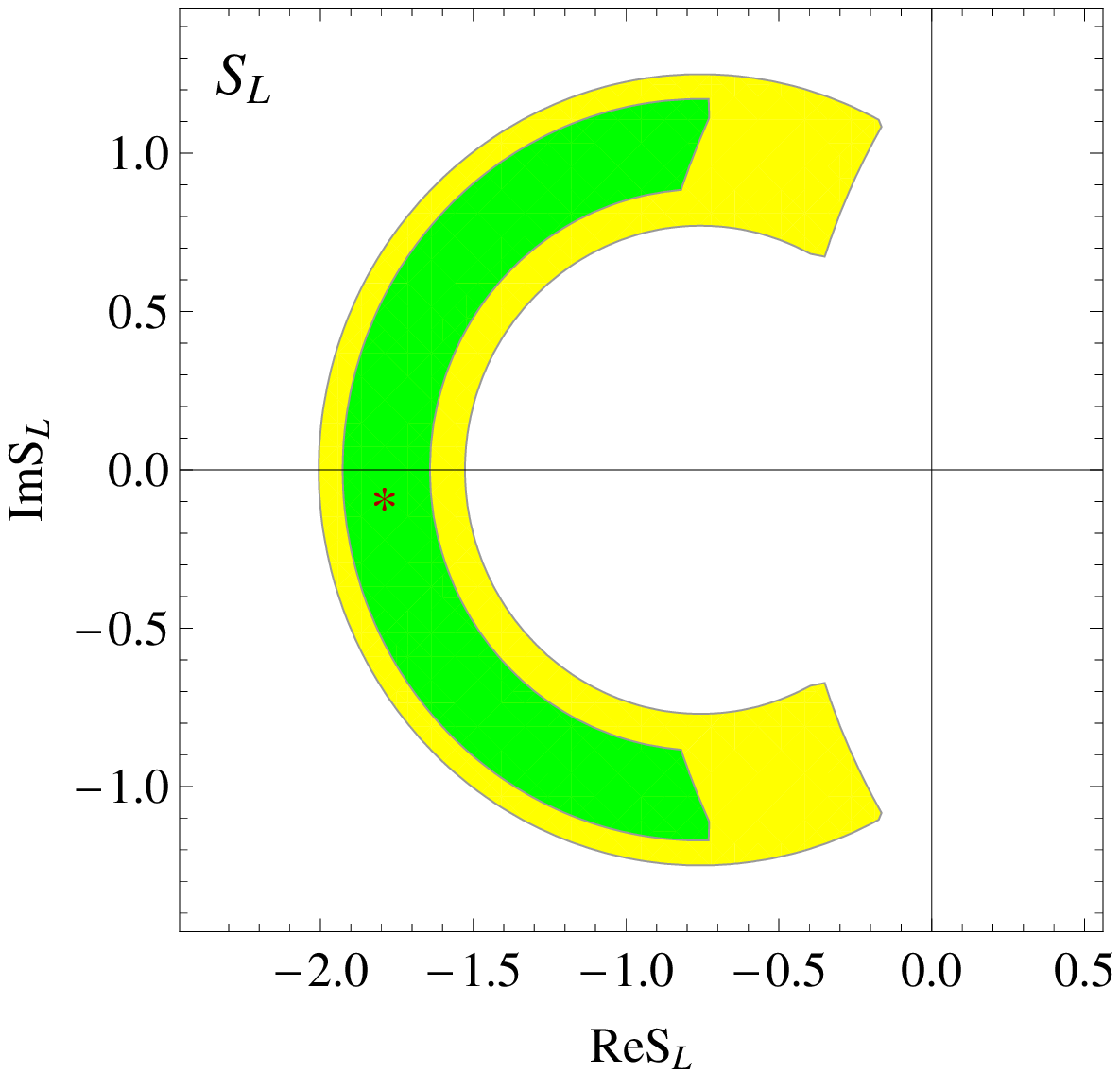}
& 
\includegraphics[scale=0.45]{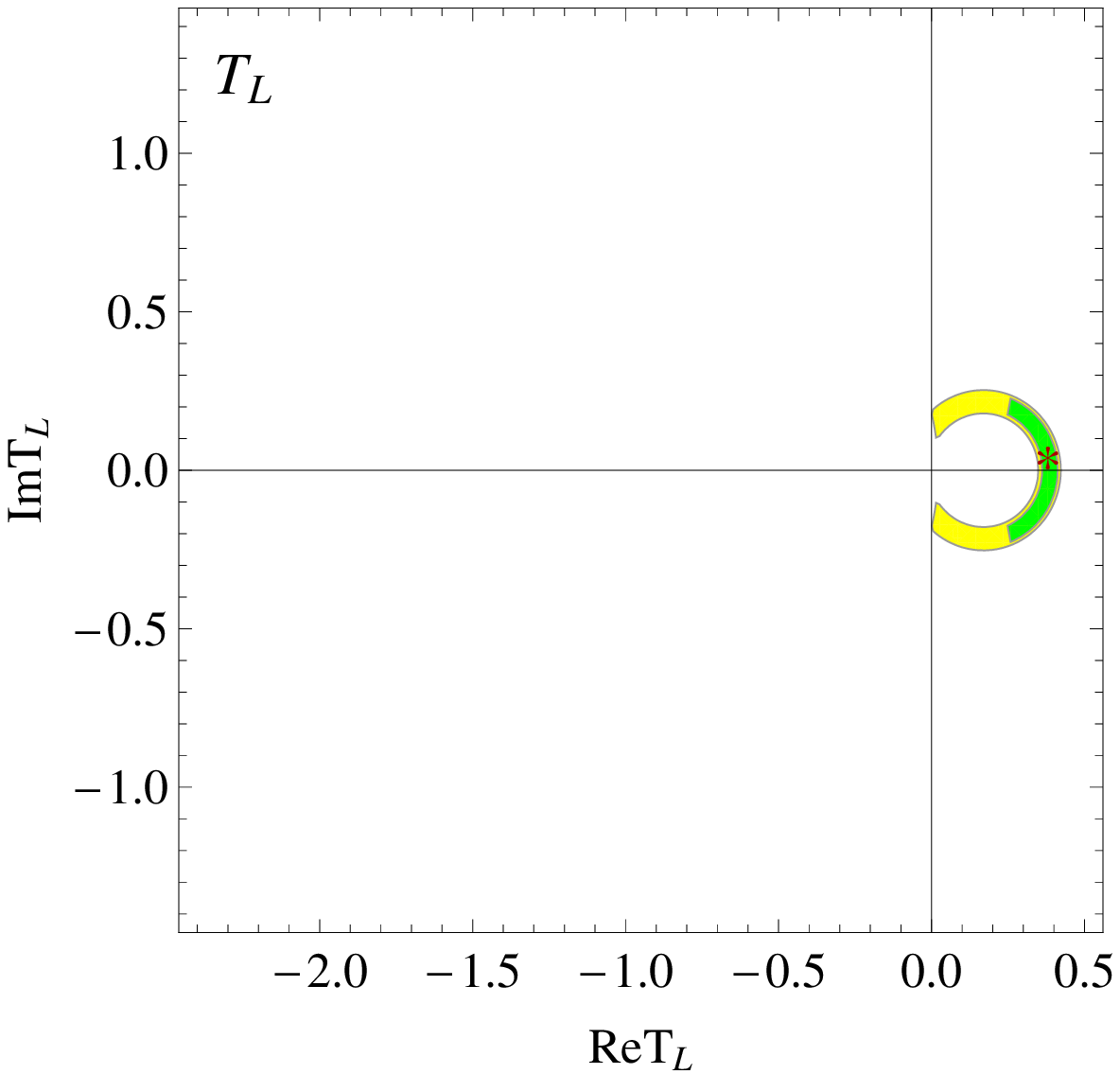}
\end{tabular}
\caption{ The allowed regions of the Wilson coefficients $V_{L,R}$, $S_L$, and 
$T_L$ within $1\sigma$ (green, dark) and $2\sigma$ (yellow, light). 
The best-fit value in each case is denoted with the symbol $\ast$. 
The coefficient $S_R$ is disfavored at $2\sigma$ and therefore  is not shown 
here.}
\label{fig:constraint}
\end{figure}

In order to acquire the allowed regions for the NP Wilson coefficients, we 
assume that besides the SM contribution, only one of the NP operators in 
Eq.~(\ref{eq:Heff}) is switched on at a time. We then compare the calculated 
ratios $R(D^{(\ast)})$ with the recent experimental data. 
The experimental constraints are shown in Fig.~\ref{fig:constraint}. 
The vector operators $\mathcal{O}_{V_{L,R}}$ and the left scalar operator 
$\mathcal{O}_{S_L}$ are favored while there is no allowed region for the right 
scalar operator $\mathcal{O}_{S_R}$ within $2\sigma$. Therefore we will not 
consider $\mathcal{O}_{S_R}$ in what follows. The tensor operator 
$\mathcal{O}_{T_L}$ is less favored, but it can still well explain the current 
experimental results. The stringent constraint on the tensor coupling mainly 
comes from  the experimental data of $R(D^\ast)$.  In each allowed region at 
$2\sigma$ we find the best-fit value for each NP coupling. The best-fit 
couplings read
\bea
V_L &=&-0.23-i0.85, \quad V_R =0.03+i0.60,\quad
S_L  = -1.80-i0.07, \quad T_L =0.38+i0.06,
\label{eq:bestfit}
\ena
and are marked with an asterisk.

The allowed regions of the coupling coefficients are then used to analyze the 
effect of the NP operators on different physical 
observables~\cite{Ivanov:2017mrj}.


\section{The decays
  $\Lambda_b \to \Lambda^{(\ast)}(\frac12^\pm,\frac32^\pm) + J/\psi$: 
matrix element  and helicity amplitudes}
\label{sec:definition}


The matrix element of the exclusive decay
$\Lambda_1(p_1,\lambda_1)\to \Lambda_2(p_2,\lambda_2)\,+\,V(q,\lambda_V)$
is defined by (in the present application the vector meson label $V$
stands for the $J/\Psi$)
\be
M(\Lambda_1\to \Lambda_2 + V) =
\frac{G_F}{\sqrt{2}} \, V_{cb} \, V^\ast_{cs} \, C_{\rm eff} \,
f_V \, M_V \, \la \Lambda_2 | \bar s O_\mu b | \Lambda_1 \ra \,
\epsilon^{\dagger\,\mu}(\lambda_V) \,,
\label{eq:matr_LbLJ}
\en
where $M_V$ and $f_V$ are the mass and the leptonic decay constant of
the vector meson $V$. The coefficient $C_{\rm eff}$ stands for the combination 
of Wilson coefficients
$ C_{\rm eff}= C_1 + C_3 + C_5+ \xi \Big(C_2 + C_4 + C_6\Big)$.
The  color factor $\xi=1/N_c$ will be set to zero such that we only keep 
the leading term in the  $1/N_c-$expansion.
The hadronic matrix element $\la \Lambda_2 | \bar s O^\mu b | \Lambda_1 \ra$
is expressed in terms of six and eight, respectively, dimensionless
invariant form factors $F^{V/A}_i(q^2)$.

The  three-quark currents with the appropriate
quantum numbers of the the $\Lambda_Q (\frac12^\pm,\frac32^\pm)$ states
are given by
\bea
\Lambda_Q^{1/2^+} &\Longrightarrow&
\epsilon_{a_1a_2a_3} \, Q_{a_1}\,
\left( u_{a_2}C\gamma_5 d_{a_3} \right),
\hspace*{1.5cm}
\Lambda_Q^{1/2^-} \Longrightarrow
\epsilon_{a_1a_2a_3} \,\gamma_5 Q_{a_1}\,
\left( u_{a_2}C\gamma_5 d_{a_3} \right) \,,
\nn
\Lambda_Q^{3/2^+} &\Longrightarrow&
\epsilon_{a_1a_2a_3} \,\gamma_5 Q_{a_1}\,
\left( u_{a_2}C\gamma_5\gamma_\mu d_{a_3} \right) \,,
\qquad
\Lambda_Q^{3/2^-} \Longrightarrow
\epsilon_{a_1a_2a_3} \,Q_{a_1}\,
\left( u_{a_2}C\gamma_5\gamma_\mu d_{a_3} \right) \,.
\label{eq:Lambda-3q-cur}
\ena
The nonlocal generalizations of the above currents are used in the CCQM
to evaluate the appropriate form factors and helicity amplitudes.

The numerical values of the normalized helicity amplitudes are listed
in Table~\ref{tab:hel}.
\begin{table}[htb]
\begin{center}
\caption{Moduli squared of normalized helicity amplitudes.}
\label{tab:hel}
\vspace*{.1cm}
\def\arraystretch{1.3}
\begin{tabular}{|c|cccc|}
\hline
$\Lambda^\ast$ & 1116 & 1405 & 1890 & 1520 \\
\hline
$J^P$         & $\frac12^+$ & $\frac12^-$ &   $\frac32^+$ &  $\frac32^-$
\\
\hline
\qquad $|\hat H_{+\tfrac32 +1}|^2$ \qquad  &\qquad 0  \qquad &\qquad 0 \qquad
&\qquad $3.50\times 10^{-4}$\qquad &\qquad $0.84\times 10^{-4}$ \qquad
\\
\qquad $|\hat H_{+\tfrac12 +1}|^2$\qquad  &\qquad $2.34\times 10^{-3}$\qquad
&\qquad $1.27\times 10^{-2}$\qquad  &\qquad $3.19\times 10^{-2}$\qquad
&\qquad $2.26\times 10^{-2}$\qquad
\\
\qquad $|\hat H_{+\tfrac12 0}|^2$\qquad   &\qquad $3.24\times 10^{-4}$\qquad
&\qquad $5.19\times 10^{-3}$\qquad  &\qquad $1.61\times 10^{-3}$\qquad
&\qquad $1.82\times 10^{-3}$\qquad
\\
\qquad $|\hat H_{-\tfrac12 0}|^2$\qquad   &\qquad 0.53 \qquad &\qquad 0.51\qquad
&\qquad $0.51$\qquad        &\qquad 0.54\qquad
\\
\qquad $|\hat H_{-\tfrac12 -1}|^2$\qquad  &\qquad 0.47 \qquad &\qquad 0.47 \qquad
&\qquad $ 0.45$ \qquad       & \qquad 0.44 \qquad
\\
\qquad $|\hat H_{-\tfrac32 -1}|^2$\qquad  &\qquad 0 \qquad &\qquad 0 \qquad
&\qquad $3.34\times 10^{-3}$\qquad  &\qquad $1.06\times 10^{-3}$ \qquad 
\\[2mm]
\hline
\end{tabular}
\end{center}
\end{table}
The helicity amplitudes $H_{\lambda_2,\lambda_V}$ of the produced $\Lambda^{(*)}$
states are clearly dominated by
the helicity configuration $\lambda_2=-1/2$ as in the quark level transition
$b \to s$. For the spin $1/2$ states in the
transition $1/2^+ \to 1/2^\pm$ this implies that the two $\Lambda^{(*)}(1/2)$
states are almost purely left-handed.

\end{document}